\documentclass[conference]{IEEEtran}
\IEEEoverridecommandlockouts
\usepackage{optidef}
\usepackage{cite}
\usepackage{booktabs}
\usepackage{amsmath,amssymb,amsfonts}
\usepackage[hang,flushmargin]{footmisc}
\usepackage{multirow}
\usepackage{float}
\usepackage{textcomp}
\usepackage{lscape}
\usepackage{url}
\usepackage{caption}
\usepackage{subcaption}

\usepackage{algpseudocode}
\usepackage{graphicx}
\usepackage{textcomp}
\usepackage{xcolor}
\usepackage{array}
\usepackage{algorithm}
\usepackage{slashbox}

\newcommand{\CASE}[1]{\STATE \textbf{case} #1\textbf{:} \begin{ALC@g}}
\newcommand{\ENDCASE}{\end{ALC@g}}

\newcommand{\DEFAULT}{\STATE \textbf{default:} \begin{ALC@g}}
\newcommand{\ENDDEFAULT}{\end{ALC@g}}
\newcommand{\DEFAULTLINE}[1]{\STATE \textbf{default:} }
\newcolumntype{L}[1]{>{\raggedright\let\newline\\\arraybackslash\hspace{0pt}}m{#1}}
\newcolumntype{C}[1]{>{\centering\let\newline\\\arraybackslash\hspace{0pt}}m{#1}}
\newcolumntype{R}[1]{>{\raggedleft\let\newline\\\arraybackslash\hspace{0pt}}m{#1}}
\def\BibTeX{{\rm B\kern-.05em{\sc i\kern-.025em b}\kern-.08em
    T\kern-.1667em\lower.7ex\hbox{E}\kern-.125emX}}

\begin{document}

\title{Sub-Graph p-cycle formation for span failures in all-Optical Networks}

\author{\IEEEauthorblockN{Varsha Lohani}
\IEEEauthorblockA{\textit{Electrical Engineering} \\
\textit{Indian Institute of Technology Kanpur}\\
Kanpur, India \\
lohani.varsha7@gmail.com}
\and
\IEEEauthorblockN{Anjali Sharma}
\IEEEauthorblockA{\textit{Electrical Engineering} \\
\textit{Indian Institute of Technology Kanpur}\\
Kanpur, India \\
anjalienix05@gmail.com}

\and
\IEEEauthorblockN{Yatindra Nath Singh}
\IEEEauthorblockA{\textit{Electrical Engineering} \\
\textit{Indian Institute of Technology Kanpur}\\
Kanpur, India \\
ynsingh@iitk.ac.in}
}

\maketitle
\begin{abstract}
p-Cycles offer ring-like switching speed and mesh-like spare capacity efficiency for protecting network against link failures. This makes them extremely efficient and effective protection technique. p-Cycles can also protect all the links in a network against simultaneous failures of multiple links. But it has been mostly studied for single link failure scenarios in the networks with the objective to minimize spare capacity under the condition of $100\%$ restorability. For large networks, use of p-cycles is difficult because their optimization requires an excessive amount of time as the number of variables in the corresponding Integer Linear Program (ILP) increase with the increase in the network size. In a real-time network situation, setting up a highly efficient protection in a short time is essential. Thus, we introduce a network sub-graphing approach, in which a network is segmented into smaller parts based on certain network attributes. Then, an optimal solution is found for each sub-graph. Finally, the solutions for all the sub-graphs is combined to get a sub-optimal solution for the whole network. We achieved better computational efficiency at the expense of marginal spare capacity increases with this approach.
\end{abstract}

\section{Introduction}

In the last few years, due to the proliferation of variety of services, including multimedia and cloud computing applications, there is enormous growth in Internet traffic. According to the Cisco Annual Internet Survey, nearly two-thirds of the world's population will have Internet connectivity by 2023. The number of nodes connected to the IP networks would be more than three times the number of people worldwide.  In 2023, average of fixed global broadband speeds will hit 110.4 Mbps, up from 45.9 Mbps in 2018\cite{cisco}. Such massive traffic volume has been feasible due to optical networks, and has been the driving force behind their further evolution.

Optical networks need to carry an enormous amount of traffic while maintaining service continuity even in the presence of faults. Failure of even a single link\footnote{span, link and edges are used interchangeably in this paper} will results in loss of a substantial amount of data if not protected automatically and restored in a very short time after the failure. Therefore, survivability against link or path failures is an essential design requirement for the high-speed optical networks. The goal of a survivability scheme is to offer reliable services for large volume of traffic even on the occurrence of failures\footnote{Fiber cut, human-made errors or natural disasters such as earthquakes, hurricanes, etc.} as well as abnormal operating conditions \cite{PR}.

p-Cycles are one of the best methods to provide protection and achieve much faster restoration speeds in case of failures. But as the network size becomes large, finding an optimum solution for p-cycle based protection becomes a time consuming task. We are trying to resolve this problem by breaking a bigger problem into smaller sub problems which can be solved in parallel on different machines. But we need to find out the best strategy of partitioning the network graph into sub-graphs with optimum size, to minimize the computation time without much increase in the spare capacity requirement (i.e., minimizing sub-optimality).

In this paper, section II discusses related research done on p-cycles and graph partitioning. In section III,  fundamental concepts of graph theory are explained. The essential understanding of graph clustering \footnote{Here, clustering and partitioning are used interchangeably} and its significance in survivability schemes has also been discussed in section III. In section IV, we present our graph partitioning methods, and in section V, we present resulting p-cycles and spare capacity requirement after partitioning. These are compared with the results for the formation of p-cycles in the whole network without any partitioning. In section VI, results for different topologies are compared to understand if the partitioning approach is a better and efficient strategy to solve the problem of finding p-cycles in a large network. Finally, the observations, future work, and conclusions are presented in the last section.

\section{p-Cycles}

Link protection schemes are used to restore the traffic through the faulty link via another path joining the two endpoints of the link \cite{link}. The other links on that primary path remain as it is. In this strategy, every link is provisioned with a backup path. The link protection schemes are fast because of faster fault localization. Two of the link protection schemes are Ring cover \cite{RC1} and p-cycles. In Ring Cover, cyclic paths are identified in the network such that each edge is traversed by at least one cycle. In this scheme, some of the links may be covered with more than one cycle, which results in additional redundancy/ required spare capacity. Therefore to minimize spare capacity, the double-cycle ring cover scheme \cite{RC2} was proposed. Each edge is equipped with protection by two unidirectional cycles. Where each cycle in the opposite direction on the shared link. The spare capacity reduces to $2N$ in this configuration. There are no cycles formed with the working capacity as done in Ring Cover protection. One of the drawbacks of double cycle ring covers is different signal delays in the forward and backward directions. Further, the restoration times are also different in the two directions.

\begin{figure}
  \centering
  \includegraphics[width=\linewidth]{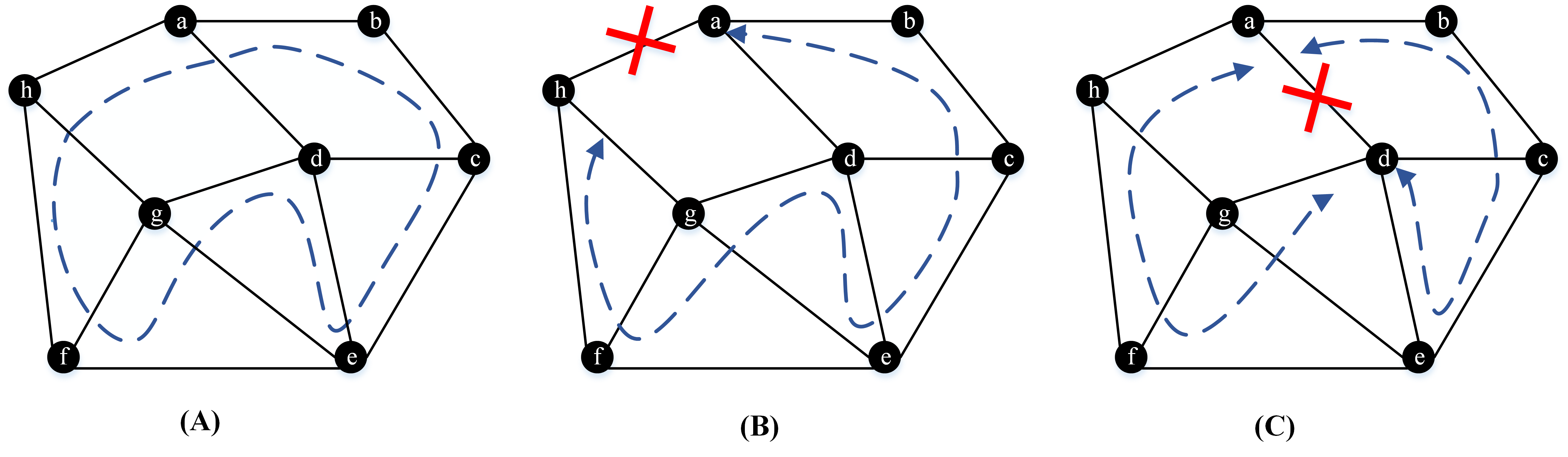}
  \caption{(A) A p-cycle, \textit{p} (\textit{a-b-c-e-d-g-f-h-a}), (B) A failed on-cycle link \textit{i} (\textit{a-h}), p-cycle \textit{p} provides single unit of backup path (\textit{a-b-c-e-d-g-f-h-a}), and (C)  A failed straddling link \textit{i} (\textit{a-d}), p-cycle \textit{p} provides two units of backup path (\textit{a-h-f-g-d}) and (\textit{a-b-c-e-d})}
  \label{fig:pc}
\end{figure}

In ring cover and double-cycle ring cover, only on-cycle failures are restored. If the straddling link\footnote{only end-nodes are part of the cycle, and not the link} failures can also be restored by the cycle, we can further reduce the spare capacity requirement. This kind of configuration is called p-Cycle based protection. The concept of p-cycles was first introduced by Grover {\em et al.} \cite{p1}. p-Cycles in an optical mesh network provide the same switching speed as the ring based protection $(< 50ms)$\footnote{Bi-directional Line Switched Ring} and capacity efficiency as in mesh networks \cite{p1}, \cite{p2}.
Whenever there is a link failure, each p-cycles based on whether the failed link is on-cycle or straddling to the p-cycle, will provide a single unit (on-cycle) or two units (straddling\footnote{A straddling link must have its edge nodes on the p-cycle, but it's not part of the cycle.}) of protection to working capacity as shown in fig. \ref{fig:pc}.

A good amount of work has been done in the past that highlights the advantages of p-cycles for the protection and restoration of traffic \cite{p1}, \cite{p2}. The p-cycles can also be used for protecting nodes \cite{node}, paths \cite{FIPP, FIPP1} and path-segments \cite{ps} in addition to links in an optical mesh network. For $100\%$ single link protection, optimum required p-cycles can be computed using an ILP (integer linear program). The main objective of the ILP is to choose the number of p-cycles to provide $100\%$ protection while minimizing the spare capacity to be used.

The concept of the Hamiltonian cycle\footnote{cycle that traverses each node once} was introduced in the literature for a single link failure scenario while all the links are carrying some traffic to be protected. It is an efficient solution when all the links are almost equally loaded, as compared to the multiple simple arbitrary p-cycles \cite{h1} based protection. It also mitigates the problem of loop-back \cite{h2}.

The p-cycle protection in large networks have also been explored using Multi-Domain approach \cite{MD1}, \cite{MD2}, \cite{MD3}, \cite{MD4}, \cite{MD5}. The idea is to partition the whole network into multiple domains using graph clustering algorithms. p-Cycles can be computed in each domain independently, forming intra-domain p-cycles. For protecting the links connecting different domains, separate inter-domain p-cycles are computed. Normally, each domain is an independently administered entity, and hence they are clearly defined by using some parameters like hop-length. The time taken by individual (small sized) domains to calculate the number of p-cycles (or spare capacity) is also expected to be less.

We would like to take clue from multi-domain protection and use partitioning of large sized optical networks to find the near optimal solutions using p-cycles in each partition independently. We can allow the node and links to be shared across two partitions to avoid inter-domain p-cycles. We will now explore how the partitions should be created and what should be their optimal sizes. One should note that this will normally be feasible only when the whole network is administered by single entity.

\section{Optical Network Graph and Clustering}

The knowledge of graph theory is imminent to appreciate the work reported in this paper. A network graph \textit{G} is defined as a set of vertices \textit{V}, indexed by \textit{v} and set of edges \textit{E}, indexed by \textit{e}. Each edge is connected to a pair of vertices e.g. \textit{$(i, j) \in E$}, where \textit{$i \in V$, $j \in V$}. 

In this paper, we represent an optical network as a graph \textit{G(V, E)}. The optical add-drop multiplexers (OADMs) in the network are the vertices, and optical fiber cables connecting these optical multiplexers are the edges. The optical edges contain the wavelength slots for WDM optical links for WDM optical networks, or frequency slots for flexi-grid (or elastic) optical networks \cite{EON1}. In this paper, we consider the undirected graphs representation of optical networks, where each span has two directed links (representing a fiber) in each direction. 

\subsection{Graph Clustering}

Graph clustering is a method of assorting the nodes of a graph into clusters such that there should be more intra-cluster links and fewer inter-cluster links \cite{gc}. Let a cluster represented by graph $G^{s}$ be composed of set of vertices $V^{s}$  ($V^{s} \subseteq V$) and set of edges $E^{s}$ ($E^{s} \subseteq E$). Therefore, the $G^{s}$ is sub-graph of $G$. Various methods of clustering a graph have been used in data science, biological and sociological networks, data transformations, database systems, etc. \cite{gc}.

In this paper, we are using clustering of the graph (we also call it graph partitioning) for resolving the problem of survivability in the optical networks with lesser computation. For large networks, graph partitioning can be very helpful in finding the near-optimal solution. Faster convergence (i.e., smaller run-time) of Integer Linear Programs \textit{(ILP)} can be achieved by running the algorithm in parallel to find a solution for all the sub-graphs.

In the literature for p-cycle based protection, the spectral clustering algorithm has been used to partition a connected graph into multi-domain sub-graphs \cite{MD3}.  In the next section, we will compare the partitioning methods of \cite{MD1}, \cite{MD2}, \cite{MD3}, \cite{MD4}, \cite{MD5} with our proposed partitioning technique.

\section{Proposed Method:Sub-Graphing using minimum cut-set algorithm}
The method proposed in this section is for both WDM based Optical Networks and Elastic Optical Networks. Here, we consider that Routing and Resource\footnote{Resource can be either Wavelength or Spectrum Slots} Allocation is done a priori. In WDM based optical network, single fixed grid slots are allocated to the connection request, which is a working demand, whereas, in EON, the set of slots can be allocated to the connection request; in this case, each request is a working demand \cite{EON2}. The set of these working demands is the working capacity as shown in table \ref{table:1}.  The method is only dependent on the working capacity traversing through the network. 

\begin{figure}
    \begin{center}
        \includegraphics[width=\linewidth]{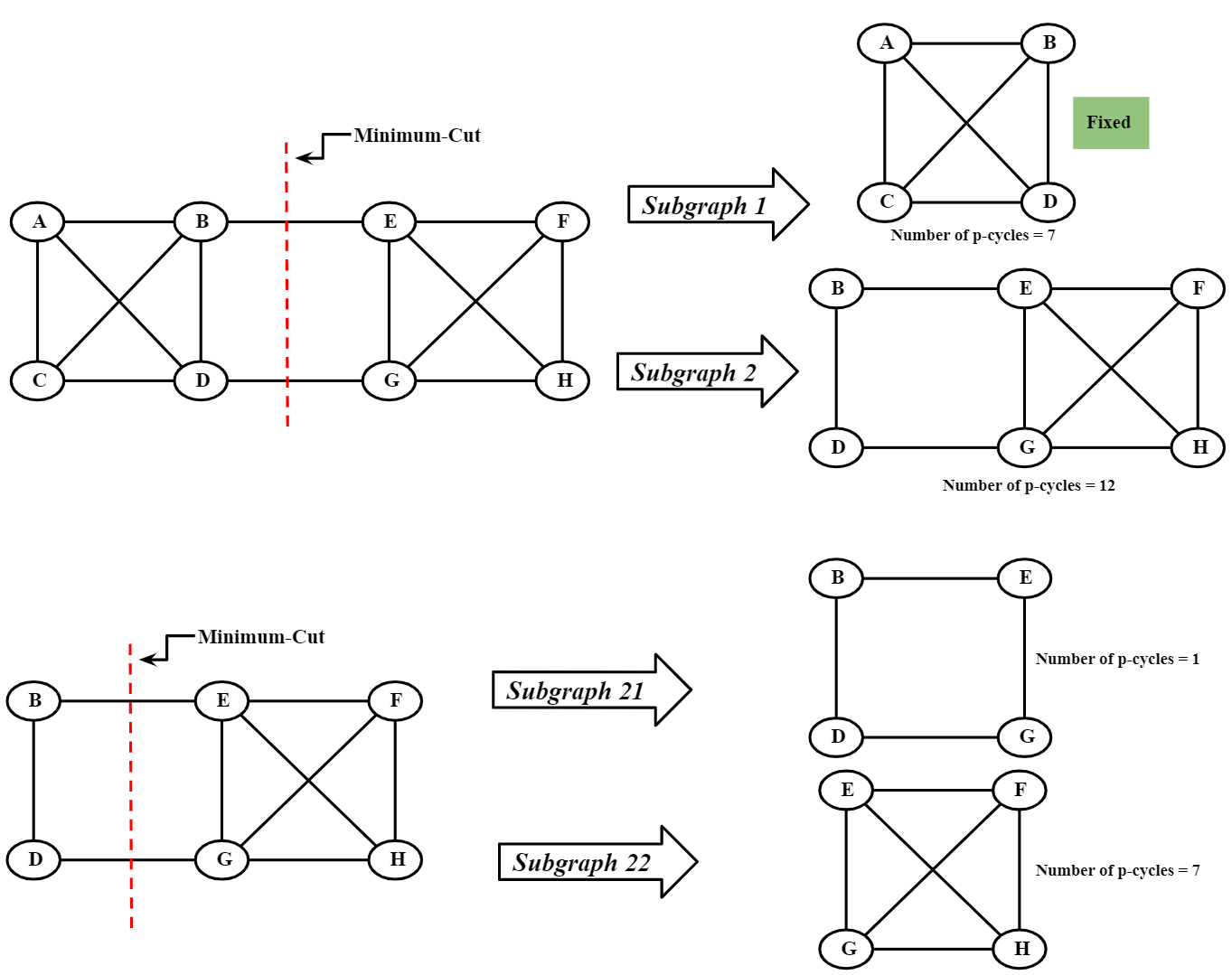}
        \caption{Sub-Graphing using minimum cut-set algorithm.}
        \label{fig:c2}
    \end{center}
\end{figure}

In this method, the entire graph is divided into two parts based on minimum cut set algorithm. The pairs of nodes which are endpoints of links in the cut-set, are used to form sub-graphs, after performing the partitioning on the super-graph using the minimum cut-set algorithm as shown in fig.\ref{fig:c2}. One of the sub-graph contains both the nodes in all such pairs. The other sub-graph contains only those nodes from the pairs, which already belong to it. Spectral partitioning method \cite{scl} is used for finding minimum cut-set as explained in Algorithm \ref{alg: SPF}. The Algorithm \ref{alg: SGF}, is used for finding the sub-graphs by further subdivisions.

Every time out of all sub-graphs, the one with most number of p-cycles is taken up for further partitioning thereby increasing the number of sub-graphs by one. The process iterates until we obtain fundamental cycles \footnote{Fundamental cycles are those cycles which does not have straddling links}.

\begin{algorithm}
\caption{Spectral Partitioning Function\textit{(SPF)}}
\label{alg: SPF}
\begin{algorithmic} [1]
\smallskip
\Function {SPF}{\textit{G(V,E)}} \Comment{\textcolor{red}{$v \in V$, $e \in E$}}

    \State {\textit{A}}=\Call{adj}{$G(V,E)$}; \Comment{\textcolor{red}{ADJ is used to find Adjacency Matrix(A) of the G(V,E)}} 
    \State {\textit{D}}=\Call{diag}{$G(V,E)$}; \Comment{\textcolor{red}{DIAG is used to find Diagonal Matrix(D) of the G(V,E)}}
    \State  \textit{L = D-A}; \Comment{\textcolor{red}{Laplacian Matrix (\textit{L})}}
    \State Compute eigenvector($ei_{v}$) of the vertices using \textit{L} matrix;
    \For {each vertices \textit{v}} 
    \If{$ei_{v}<0$}
    \State place vertix $v$ to $V^{(1)}$
    \Else
    \State place vertix $v$ to $V^{(2)}$
    \EndIf
    \EndFor
    \State Using $V^{(1)}$ and $V^{(2)}$ find Minimum cut-set;
    \State Mark Minimum cut-set as \textbf{\textit{Slider}};
    \State \Return $g^{(1)}$, $g^{(2)}$ \Comment{\textcolor{red}{$g^{(i)}$ = $g^{(i)}(V^{(i)}, E^{(i)})$ where i = 1,2.}}
\EndFunction
\smallskip
\end{algorithmic}
\end{algorithm}

\begin{algorithm}
\caption{Function for Sub-Graphing using p-cycle}
\label{alg: SGF}
\begin{algorithmic} [1]
\smallskip
\Function {PCycleSub}{${G(V,E), P, P_{s}}$} 
    \State Call [$g^{(1)}$, $g^{(2)}$] = SPF(\textit{G(V,E)})
    \State $p_{g^{(1)}}$ = \Call{PCycle}{$g^{(1)}$};
    \State $p_{g^{(2)}}$ = \Call{PCycle}{$g^{(2)}$};
    
        \If{$p_{g^{(1)}}>=1$ $\&\&$ $p_{g^{(2)}}>=1$} 
            
		    \State $P_{s}$.\Call{Append}{[${g^{(1)}}, p_{g^{(1)}}$]};
		     \State $P_{s}$.\Call{Append}{[${g^{(2)}}, p_{g^{(2)}}$]};
		    \State $Sum_p$ = null;
		    \State $Sum_p$ = \Call {Sum}{$P_{s}$}; \Comment{\textcolor{red}{SUM function is to find out the sum of pcycles (not graph)}}
		    \State \textit{P}.\Call{Append}{$Sum_p$};
		    \State {$maxP_s$} = \Call{Max}{$P_{s}$}; \Comment{\textcolor{red}{MAX function is to find out the element([graph, pcycle]) with maximum $p_{g^{()}}$ value from $P_s$}}
		    \State $P_{s}$.\Call{Remove}{\textit{max$P_s$}};
        \EndIf
        \State \Call{PCycleSub}{$maxP_s[0]$, $P, P_{s}$} \Comment{\textcolor{red}{input argument takes sub-graph with highest number of p-cycles}}

\EndFunction
\smallskip
\end{algorithmic}
\end{algorithm}

\begin{algorithm}
\caption{Sub-Graphing Algorithm}
\label{alg: SGA}
\begin{algorithmic} [1]
\smallskip
\State $p_{org}$ = \Call{PCycle}{\textit{G(V,E)}}; \Comment{\textcolor{red}{original graph p-cycle count}}
\State \textit{P} = [$p_{org}$]; \Comment{\textcolor{red}{p-cycle count for each partition}}
\State $P_{s}$ = [];  \Comment{\textcolor{red}{contains sub-graphs p-cycle count}} 
\State \Call {PCycleSub}{$G(V,E),P, P_{s}$}
\smallskip
\end{algorithmic}
\end{algorithm}

In algorithms \ref{alg: SGF} and \ref{alg: SGA}, PCYCLE() function is used to calculate the number of p-cycles within a graph. Algorithm \ref{alg: SGF} is a recursive function used for the sub-graphing of a graph based on the p-cycles count. In each recursion, the sub-graphs with the maximum number of p-cycles is separated from the list and further partitioned into two using Spectral Partitioning Function(SPF). The partitioning continues until all the sub-graphs are simply fundamental p-cycles. Algorithm \ref{alg: SGA} gives the list of p-cycles for every partition.  This list is used in subsequent section for spare capacity optimization and compute time analysis. 

Intuitively, as the number of partitions increases, the size of the sub-graphs reduces. The run time for optimizing the p-cycle is reduced, considering that the number of candidate p-cycles and the number of all other variables decrease. The optimizations for all the partitions are assumed to be done in parallel. However, the spare capacity requirement will increase. Nevertheless, for real-time scenarios, a much faster computation of protection and hence provisioning will be possible.

\subsection{ILP Optimization for Sub-graphs}

The Sub-Graphing Algorithm \ref{alg: SGA} returns the partitioned graph with the p-cycle count for each partition. In this section, with the help of Integer Linear Programming (ILP), we are calculating spare capacity requirement and analyzing compute time for each partition. The set of candidate p-cycles are used as an input to the ILP.

The ILP is used here for spare capacity optimization for $100\%$ protection against the single link failure in the network sub-graph. It may be noted that as optimization is done for each sub-graph, and therefore failure happening in different sub-graphs simultaneously, can be protected independently. The objective is to minimize the total spare capacity \cite{p1} for each sub-graph, $G^{s}$. In this approach, all candidate p-cycles for each sub-graph are pre-computed.\\

\noindent
\textbf{SETS}\\
\begin{tabular}{p{0.2cm}p{0.1cm}p{6.5cm}}
$L^{s}$ & : & Set of link in sub-graph \textit{s}, indexed by \textit{l}.\\
$P^{s}$ & : & Set of p-cycles in sub-graph \textit{s}, indexed by \textit{p}.
\end{tabular}\\
\\

\noindent
\textbf{PARAMETERS}\\
\begin{tabular}{p{0.2cm}p{0.1cm}p{6.5cm}}
$C^{s}_{l}$ & : & Cost of link \textit{l} in sub-graph \textit{s},\\
$W^{s}_{l}$ & : & Working capacity on link \textit{l} in sub-graph \textit{s},\\
$N^{s}_{p}$ & :	& Number of p-cycles \textit{p} in sub-graph \textit{s},\\
$S^{s}_{l}$ & : & Spare capacity on link \textit{l} in sub-graph \textit{s},\\
\end{tabular}
\vspace{2mm}
\noindent
\begin{equation*}
    \delta_{l}^{p} = \begin{cases}
    1, &\text{if p-cycle \textit{p} crosses link \textit{l},}\\
    0, &\text{otherwise,}
\end{cases}
\vspace{2mm}
\end{equation*}
\begin{equation*}
    x_{l}^{p} = \begin{cases}
    1, &\text{if p-cycle \textit{p} protects link \textit{l} as on-cycle span,}\\
    2, &\text{if p-cycle \textit{p} protects link \textit{l} as straddling span,}\\
    0, &\text{otherwise,}
\end{cases}
\end{equation*}

\textit{Single Link Failure}- The objective is to minimize the spare capacity which is used to form the p-cycles in the sub-graph to protect all possible single link failures.\\
\\
\textbf{\textit{Objective:}} \textit{Minimize} $\biggl(\sum_{l \in L} C^s_{l} * S^s_{l} \biggr)$\\
\\
\textbf{\textit{Subject to:}}
\begin{itemize}
    \item All the working capacity of every link ( $100\%$ protection) is protected against any possible single link failure.
    \begin{equation*}
    W^{s}_{l} \leq {\sum_{p \in P^{s}} x_{l}^{p}* N^{s}_{p}}, \hspace{0.5in} \forall ~ l \in  L^{s}
    \end{equation*}
    
    \item Enough spare capacity exists on each link to form the p-cycles.
    \begin{equation*}
    S^{s}_{l} \geq {\sum_{p \in P} \delta_{l}^{p} * N^{s}_{p}},  \hspace{0.5in} \forall ~ l \in  L^{s}
    \end{equation*}
    
    \item $N^{s}_{p} \geq 0$.\\
\end{itemize}

The above ILP assumes that all nodes have full wavelength conversion capability.

\section{Results for different topology}

\begin{figure}
\begin{subfigure}{.22\textwidth}
  \centering
        \includegraphics[width=\linewidth]{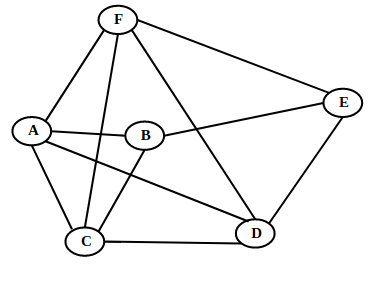}
        \caption{Net1: N06L11}
        \label{fig:net1}
\end{subfigure}
\begin{subfigure}{.22\textwidth}
  \centering
  \includegraphics[width=\linewidth]{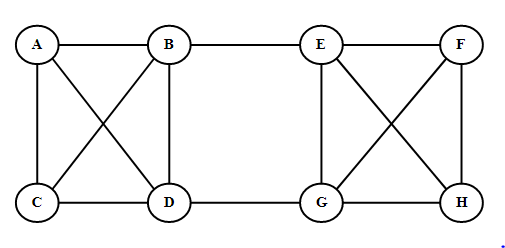}
        \caption{Net2: N8L14}
        \label{fig:net2}
\end{subfigure}
\begin{subfigure}{.22\textwidth}
  \centering
  \includegraphics[width=\linewidth]{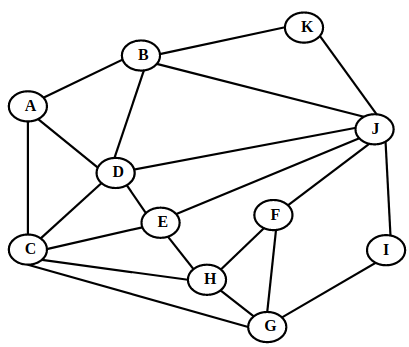}
        \caption{Net3: COST239}
        \label{fig:net3}
\end{subfigure}
\begin{subfigure}{.22\textwidth}
  \centering
        \includegraphics[width=\linewidth]{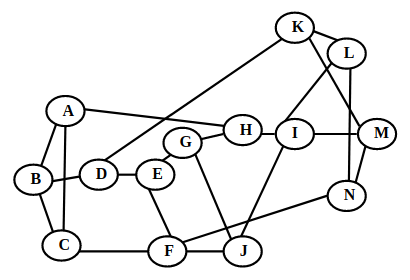}
        \caption{Net4: NSFNET}
        \label{fig:net4}
\end{subfigure}
\caption{Different Network Topologies used for Simulation and Optimization.}
\label{fig:fig}
\end{figure}

\begin{figure}
    \begin{center}     
        \includegraphics[width=\linewidth]{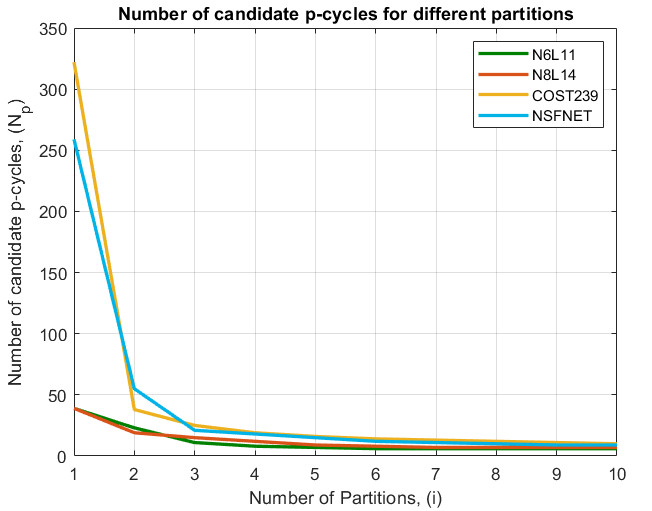}
        \caption{Number of Candidate p-cycles Vs Number of Partitions for different network topologies.}
        \label{fig:r4}
    \end{center}
\end{figure}

\begin{figure}
    \begin{center}                    
        \includegraphics[width=\linewidth]{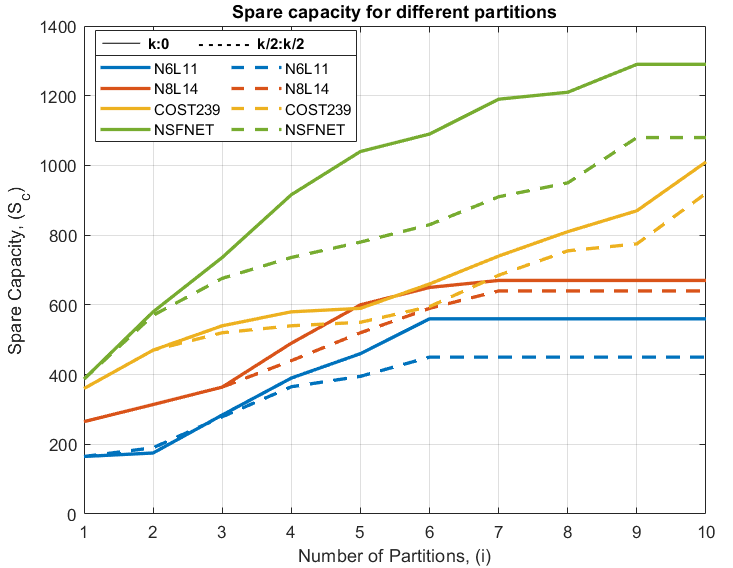}
        \caption{Spare Capacity Vs Number of Partitions for different network topologies}
        \label{fig:r3}
    \end{center}
\end{figure}

\begin{figure}
    \begin{center}                    
        \includegraphics[width=\linewidth]{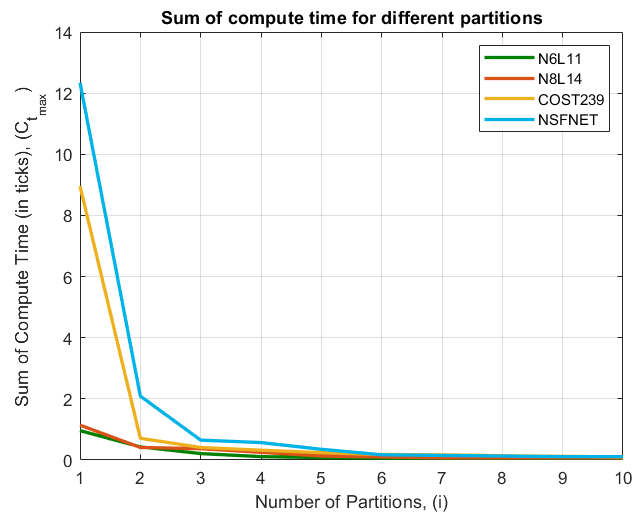}
        \caption{Sum of Compute Time Vs Number of Partitions for different network topologies.}
        \label{fig:r2}
    \end{center}
\end{figure}

\begin{figure}
    \begin{center}
        \includegraphics[width=\linewidth]{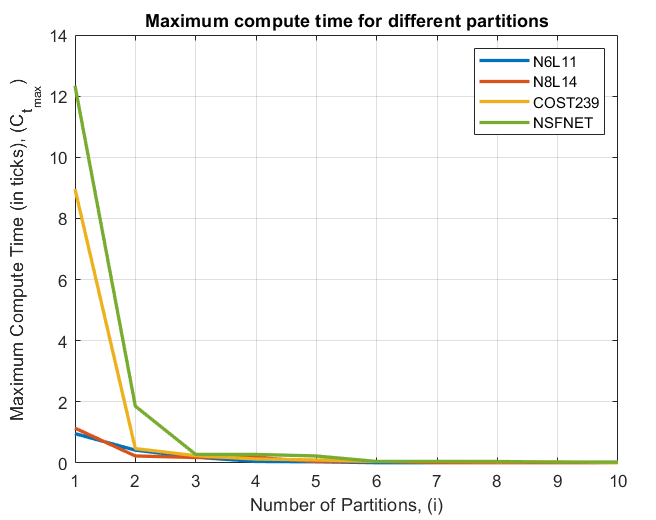}
        \caption{Maximum Compute Time Vs Number of Partitions for different network topologies.}
        \label{fig:r1}
    \end{center}
\end{figure}

We consider the four network topologies as shown in figs.\ref{fig:net1}, \ref{fig:net2}, \ref{fig:net3}, and \ref{fig:net4}. SCILAB 5.5.2, and CPLEX 12.9.0, are used for network parameters calculation and optimization respectively. Table \ref{table:1} displays the amount of total working capacity assigned to all links in each network. For different partitions of network topologies, different attributes (Number of p-Cycles, Spare Capacity, and Compute Time) are computed and plottted in figs. \ref{fig:r4}, \ref{fig:r3}, \ref{fig:r2} and \ref{fig:r1}. We estimated two compute times- the overall computation time in fig. \ref{fig:r2} and the maximum computation time in fig. \ref{fig:r1}. The overall computation time is the sum of all sub-graphs' computation time in the $i^{th}$ iteration of partitioning. If we run the optimization on all the sub-graphs in parallel, then the maximum computation time is selected from the times taken by each of them.

The working capacity distribution between the common links of the two partitions is $0:k$ or $k:0$ (Type I) based on spare capacity minimization, where $k$ is the working capacity on the common link. But the spare capacity for this sort of distribution turns out to be very high. So, we chose equal distribution of working capacity on the common link to the two partitions, i.e. $k/2 : k/2$ (Type II). As we can see from fig. \ref{fig:r3} the value of spare capacity is high when less number of partitions are made for both Type I and Type II working capacity distributions. But with the increasing number of partitions, the spare capacity requirement reduces for Type II. So, opting for $k/2 : k/2$ working capacity distribution is better. The dynamic partitioning is also possible $k/i : k/j$\footnote{\textit{i} and \textit{j} are the whole numbers.} but the results are almost same as $k/2 : k/2$ for less partitions. Since the spare capacity values are much higher for more partitions, we are ignoring them.

Here, we are partitioning the networks to get faster configuration of protection, which is desirable for large networks as well as real-time scenarios. As we can see in figs.\ref{fig:r4}, \ref{fig:r2}, and \ref{fig:r1}, that after fourth partitioning iteration the results are almost constant for all the network topologies. But, here we seek to keep partitioning as minimal as possible so that the spare capacity used is not high and the remaining capacity can be used for another failed link which is disjoint to the protection of first failed link. We can see in fig.\ref{fig:r4}, that the number of candidate p-cycles are significantly reduced for all the network topologies. As the candidate p-cycles are reduced, the number of variables in the ILP and hence the compute time also decreases. So, we can go for partitions starting from 2 until the results are almost constant (i.e., the near-optimal partition, in our case four partitions).

\begin{table*}[ht]
    \begin{center}
        \centering
        \caption{Assigned capacities for various network topologies}
        \label{table:1}
        \begin{tabular}{|c|c|c|c|}
            \hline
            \centering
            \textbf{Networks} & \textbf{Total Capacity} & \textbf{Working Capacity}& \textbf{Maximum allowable Spare Capacity} \\
            \hline
            Net1 &             880  &              300&            580\\
            \hline
            Net2 &              1120 &              390&           730\\
            \hline
            Net3 &              1680 &              510&            1170\\
            \hline
            Net4 &              1760 &              550&            1210\\
            \hline 
        \end{tabular}
    \end{center}
\end{table*}

Sub-graphing method reduces the compute time for finding p-cycle based protection configuration in  real-time scenario. p-Cycles have not found practical use only due to large compute time for finding optimal strategy for big networks. But now, the protection configuration can be determined quickly as a consequence of sub-graphing.  The partitioning methods used \cite{MD1}, \cite{MD2}, \cite{MD3}, \cite{MD4}, \cite{MD5} considers inter-domain and intra-domain. Each intra-domain is provisioned with different sets of p-cycle whereas for inter-domain either FIPP p-cycles are used or different approach is used to provision protection. This increases the compute time for ILP.

Another alternative approach in real time protection is to do local information based decision using heuristics at all nodes. Our algorithm can be compared with such actual real-time scenarios where the protection is computed dynamically on need basis. In such dynamic algorithms, when a p-cycle is not protecting any working path, it should be dismantled to recover free capacity from other p-cycles on need basis. The existing p-cycles should be used whenever possible. It is possible that we can set up a working path but can't protect all of its links. These p-cycles can be merged or expanded to keep the protection capacity to a minimum in such a scheme. It may not always work with optimum capacity, but can be pushed towards optimality.

By partitioning, we can now periodically compute the optimal configuration and correct the operating p-cycles. Thus, the dynamic creation, dismantling, merger and expansion of p-cycles is not needed which will always give sub-optimal performance. But the ILP for the partition should consider the total fixed capacity in all the links. The objective will change to provide maximum protection, and consequently sometimes the optimal solution will not provide $100 \%$ single fault tolerance.

For real-time scenario, we can write heuristics where we can assign p-cycles for protection without interrupting existing traffic on all other routes, though there is no guarantee of  $100 \% $ restorability. However, we can try to merge p-cycles of different partitions and get optimal partitioning until we get enough spare capacity to protect the failed links to the extent possible.

\section{Conclusion}

We have formulated sub-graphing method that splits the network graph into several partitions, so that optimization for each partition can be made in parallel. But, with this, the key issue is the increase in the amount of spare capacity needed. So, we have tried to find out near-optimal partitions with which the both the requirement of minimum compute time and spare capacity can be met. The results indicate that this method hold promise in making p-cycles practically feasible.

\end{document}